\documentclass[a4paper,11pt]{article}
\usepackage{pos}

\def\slashchar#1{\setbox0=\hbox{$#1$}
	\dimen0=\wd0
	\setbox1=\hbox{/} \dimen1=\wd1
	\ifdim\dimen0>\dimen1
	\rlap{\hbox to \dimen0{\hfil/\hfil}}
	#1
	\else
	\rlap{\hbox to \dimen1{\hfil$#1$\hfil}}
	/
	\fi}

\newcommand{\FIG}{Fig.~}

\newcommand{\SEC}{Sec.~}

\newcommand{\EQ}{Eq.~}

\newcommand{\GeV}{\,\mbox{GeV}}

\def\bei{\begin{itemize}}
	\def\ei{\end{itemize}}

\def\beeq{\begin{eqnarray}} 
	\def\beqa{\begin{eqnarray}}
		\def\bea{\begin{eqnarray}}
			
			\def\eea{\end{eqnarray}}
		\def\eqa{\end{eqnarray}}
	\def\eeeq{\end{eqnarray}}

\def\eqar{\end{array}}
\def\beqar{\begin{array}}

\def\beas{\begin{eqnarray*}}
	\def\beqas{\begin{eqnarray*}}
		
		\def\eqas{\end{eqnarray*}}
	\def\eeas{\end{eqnarray*}}

\def\beq{\begin{equation}} 
	\def\be{\begin{equation}}
		
		\def\ee{\end{equation}}
	\def\eq{\end{equation}}
\def\eeq{\end{equation}}

\def\beqd{\begin{displaymath}}
\def\eeqd{\end{displaymath}}
\def\eqd{\end{displaymath}}

\def\beeq{\begin{eqnarray}} \def\eeeq{\end{eqnarray}}

\newcommand{\fin}{\end{document}}

\def\fin{\end{document}}

\newcommand{\REF}{Ref.~}

\title{Complete 1-loop study of exclusive $ J/\psi $ and $ \Upsilon $ photoproduction  with full GPD evolution}
\ShortTitle{Complete 1-loop study of exclusive $ J/\psi $ and $ \Upsilon $ photoproduction}

\author[a]{Chris Flett}
\author[a]{Jean-Philippe Lansberg}
\author*[a,b]{Saad Nabeebaccus}
\author[a]{Maxim Nefedov}
\author[c]{Pawel Sznajder}
\author[c]{Jakub Wagner}

\affiliation[a]{Universit\'e Paris-Saclay, CNRS, IJCLab, 91405 Orsay, France}

\affiliation[b]{Department of Physics and Astronomy, University of Manchester, Manchester, M13 9PL, United
	Kingdom}

\affiliation[c]{National Centre for Nuclear Research (NCBJ), Pasteura 7, 02-093 Warsaw, Poland}

\emailAdd{Christopher.Flett@ijclab.in2p3.fr}
\emailAdd{Jean-Philippe.Lansberg@ijclab.in2p3.fr}
\emailAdd{saad.nabeebaccus@manchester.ac.uk}
\emailAdd{Maxim.Nefedov@ijclab.in2p3.fr}
\emailAdd{Pawel.Sznajder@ncbj.gov.pl}
\emailAdd{Jakub.Wagner@ncbj.gov.pl}

\abstract{We discuss the exclusive photoproduction of a heavy vector quarkonium, namely $J/\psi$ and $\Upsilon$ at 1-loop in $\alpha_s$. In collinear factorisation (CF), the amplitude for such a process is obtained by the convolution of a hard partonic sub-amplitude, with a universal generalised parton distribution (GPD). For the first time, we perform a complete calculation at 1-loop including full leading-log (LL) GPD evolution. We first demonstrate the huge instability of the cross section at high energies when the factorisation scale $\mu_F$ is varied. This instability has been reported previously in the literature, and occurs due to the large logarithms generated by the huge difference between the hard scale of the process, which is the mass of the heavy quarkonium here, and the centre-of-mass energy of the process. This problem was also reported in \textit{inclusive} heavy vector quarkonium photoproduction. We show that this issue can be resolved by resumming these large logarithms using high-energy factorisation (HEF), by performing a matching with the result in CF using the doubly logarithmic approximation (DLA) in order to be consistent with the fixed order evolution of GPDs. Finally, we show that the cross sections obtained from such a matching, besides being free from the previously mentioned factorisation scale variation instabilities, are consistent with the H1 data for $J/\psi$ production and with the ZEUS data for $\Upsilon$ production.}

\FullConference{31st International Workshop on Deep Inelastic Scattering (DIS2024)\\
 8–12 April 2024\\
Grenoble, France\\}

\begin{document}
\maketitle

\section{Introduction}

\begin{figure}
	\centering
	\raisebox{0.5cm}{\includegraphics[width=7.5cm]{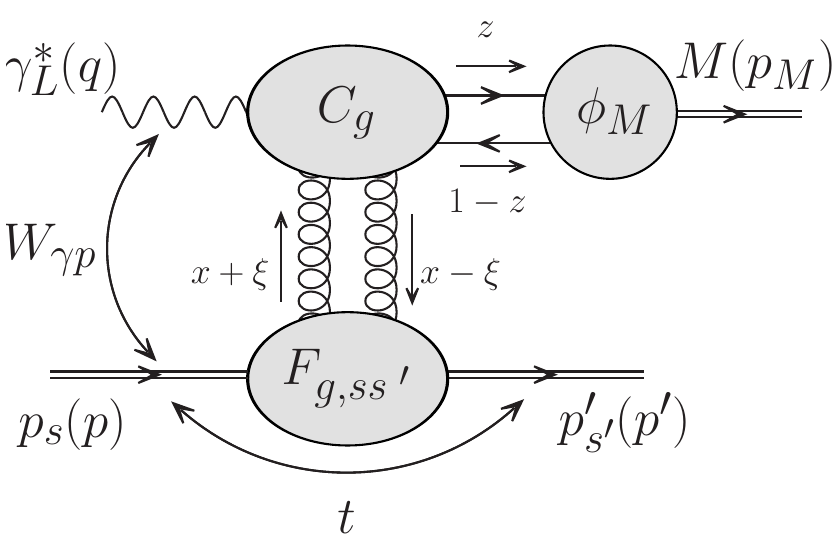}}
	\hfill
	\includegraphics[width=5.5cm]{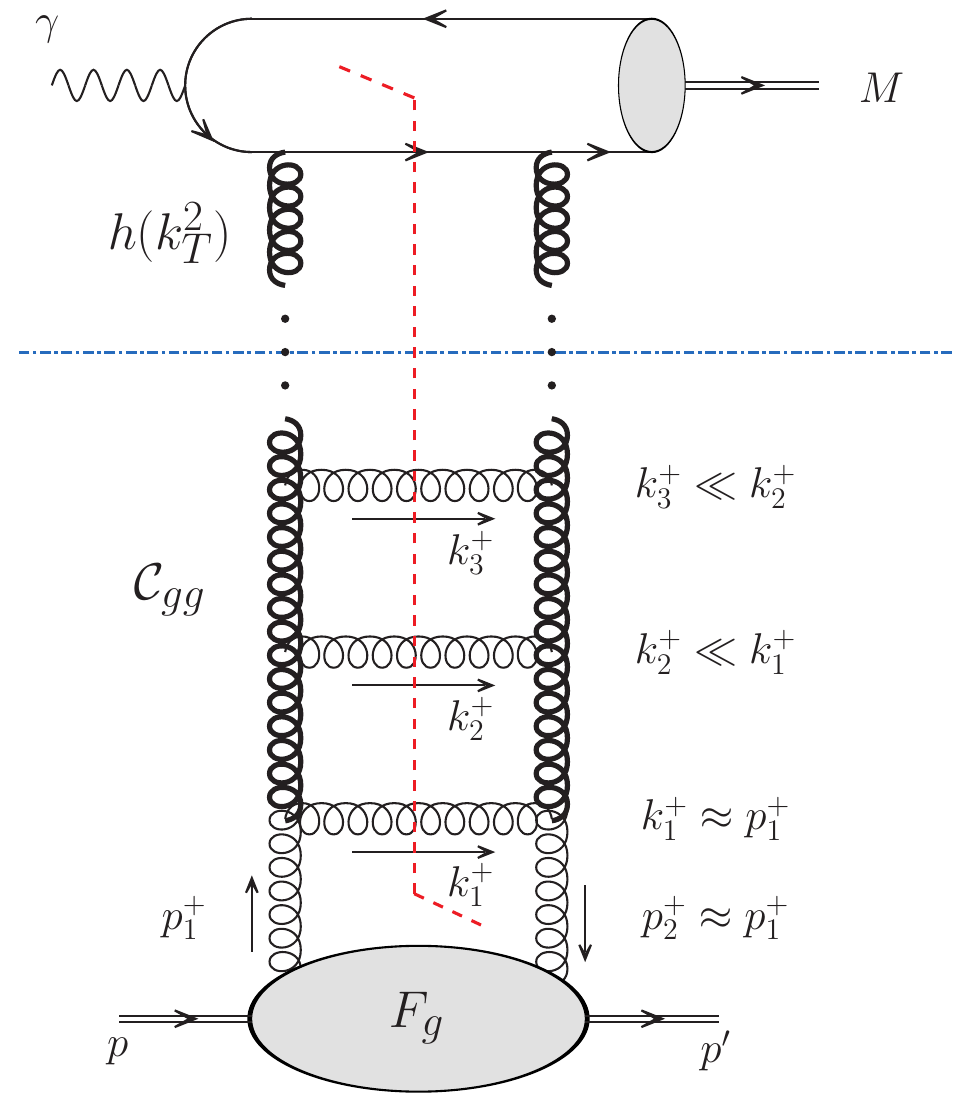}
	\caption{Left: Factorised amplitude for DVMP. Right: An example Feynman diagram that contributes to the imaginary part of the amplitude in \EQ\eqref{eq:amp} in the $ \rho \ll 1 $ region of the LLA. Reggeised gluons are shown in bold. See the text for further explanation of the notation.}
	\label{fig:dvmp}
\end{figure}

Deeply-virtual meson production (DVMP) is an important process that can be studied in collinear factorisation (CF) in order to probe \textit{generalised parton distributions} (GPDs) \cite{Collins:1996fb}. DVMP on a proton can be represented as $\gamma^{*}_{L}(q) + p_{s}(p) \to M(p_M) p'_{s'} (p')$, where the subscript $L$ stands for longitudinally polarised, and $s$ and $s'$ are the proton spins in the initial and final state respectively. A key feature of CF for exclusive processes is that it occurs at the amplitude level. For DVMP, one has
\begin{align}
	{\cal M}_{ss'} = \sum_{i = q,g}\int_{-1}^{1} dx \int_{0}^{1}dz\,C_{i}(x,\xi,z;\mu_{F},\mu_{R})\,F_{i,\,ss'}(x,\xi,t;\mu_{F})\, \phi_M(z;\mu_{F})\,.
\end{align}
Diagrammatically, this is shown on the left panel of {\FIG\ref{fig:dvmp}}. The factorised amplitude $ {\cal M}_{ss'} $  for DVMP involves the convolution of three objects: (i) the hard partonic sub-process, or coefficient function, $ C_{i}(x,\xi,z;\mu_{F},\mu_{R}) $, which can be calculated perturbatively in QCD, (ii) a GPD $F_{i,\,ss'}(x,\xi,t;\mu_{F})$, and (iii) a \textit{distribution amplitude} (DA) $ \phi_{M}(z;\mu_{F}) $, which is a universal non-perturbative function describing the transition between a collinear $q\bar{q}$ pair  to a meson state. In the above, $x$ and $z$ are momentum fractions of the collinear partons linking the coefficient function to the GPD and DA respectively, while the index $i = q,g$ is the type of parton (quarks or gluons) probed from the proton. $ \mu_{F} $ and $ \mu_{R} $ are the factorisation and renormalisation scales respectively, and $ \xi $ is the so-called \textit{skewness} parameter, which describes the net transfer of longitudinal momentum from the nucleon sector to the hard sub-process,\footnote{We use \textit{light-cone coordinates}, such that for a generic 4-vector $ k $, $ k^{\pm} = (k^0 \pm k^3)/\sqrt{2} $ and $ \mathbf{k}_{\perp} = (k^1,k^2) $. Throughout these proceedings, we work in the centre-of-mass frame of the $\gamma p$ system, with the $ z $-axis aligned along the proton direction.} $ \xi = \Delta^+/(p+p')^{+} $, and $\Delta = p' - p$ with $t = \Delta^2$. The GPD, being a function of $ x,\xi$ and $ t $, provide 3D information on the distribution of partons inside nucleons. In fact, GPDs reduce to the usual 1D parton distribution functions (PDFs) in the forward limit, given by $ \xi,t \to 0 $.

Going beyond DVMP, we consider here the \textit{photoproduction} of a heavy vector quarkonium, using the heavy quark mass as the hard scale to justify the collinear factorisation in that case. While there is no formal proof of factorisation to date, the 1-loop computation of \REF\cite{Ivanov:2004vd} shows that it holds at next-to-leading order (NLO) in  $ \alpha_s $. The calculation involves the so-called \textit{static limit}, where the DA of the outgoing heavy meson $ \phi_M (z) $ reduces to $\delta (z-1/2)$. This calculation has been extended to electroproduction in \REF\cite{Flett:2021ghh}.

\section{Amplitude in collinear factorisation}

\label{sec:amp-coll-fact}

The amplitude for vector quarkonium photoproduction can be written as
\begin{align}
	\label{eq:amp}
{\cal M}_{ss'}^{ \lambda  \lambda '} = - \varepsilon_{ \lambda }^{\gamma }\cdot  \varepsilon _{ \lambda '}^{*{Q}} \sum_{i=q,g}\int_{-1}^{1} \frac{dx}{x^{1+ \delta _{ig}}}C_{i}(x, \xi ; \mu _{F}, \mu _{R}) F_{i,\,ss'}(x, \xi ,t; \mu _{F})\,,
\end{align}
where $  \varepsilon _{ \lambda }^{\gamma } $ and $  \varepsilon _{ \lambda '}^{* Q} $ are the polarisation vectors of the incoming photon and outgoing quarkonium respectively.  An important and interesting feature of heavy quarkonium production is that, at LO in $ \alpha_s $, the amplitude is sensitive to gluon GPDs only. Sensitivity to quark GPDs arises at NLO.

At LO in $ \alpha_{s} $, and in the relative velocity of the heavy quarks, we have
\begin{align}
	C_{g}^{(0)}(x, \xi ;\mu_{R}) = \frac{x^2 c}{(x+\xi-i\epsilon )(x-\xi + i \epsilon )}\,,
\end{align}
where $ c =  (4 \pi  \alpha _{s}( \mu _{R})ee_{Q}R_{Q}(0)/(m_{Q}^{\frac{3}{2}}\sqrt{2 \pi N_{c}}))$. In the above, $ e_{Q} $ is the charge of the heavy quark in units of $ e $, which is the charge of the positron in natural units. $ R_{Q}(0) $ is the value of the spatial radial wave function at the origin, and $ m_{Q} = M_{Q}/2 $ is the heavy quark mass, taken to be half the mass of the quarkonium, $ M_{Q} $.

\begin{figure}
	\includegraphics[width=7.5cm]{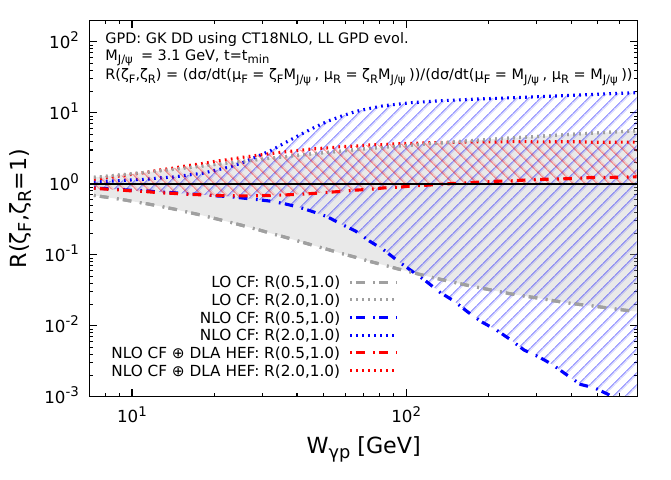}
	\hfill
	\includegraphics[width=7.5cm]{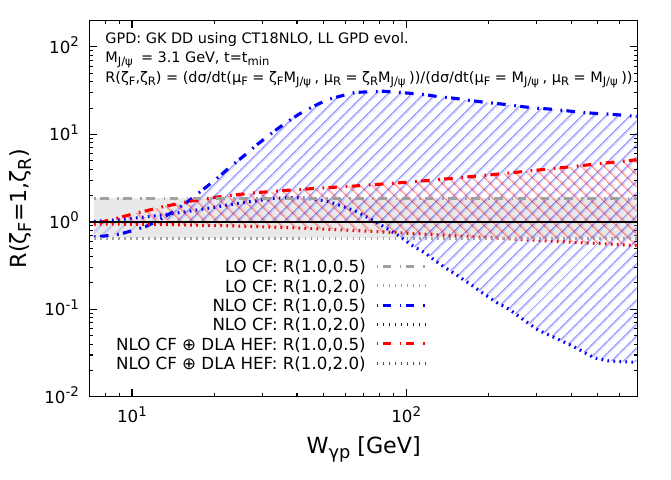}
	\caption{Left: $  \mu _{F} $ uncertainty for $ J/\psi $ photoproduction using coefficient functions from LO CF (grey), NLO CF (blue) and matched CF and HEF (red). Right: The same but for the $  \mu _{R} $ uncertainty.}
	\label{fig:jpsi-uncertainty}
\end{figure}

While the NLO coefficient is too lengthy to be displayed here, it is nevertheless instructive to study its limit at high centre-of-mass energies $ W_{\gamma p} $, or equivalently small $ \xi \approxeq M_{Q}^2/(2 W_{\gamma p})  $. One finds that the NLO coefficient functions $ C_{i}^{(1)}(x,\xi;\mu_{F}, \mu _{R}) $  scale as
\begin{align}
	\label{eq:asy-amp}
-\frac{i \pi c |x|}{2 \xi } \frac{ \alpha _{s}( \mu _{R})}{ \pi }\ln  \left( \frac{m_{Q}^2}{ \mu _{F}^2} \right) \{C_{A},\,2C_{F}\} \equiv C^{(1,\, \mathrm{asy} .)}_{\{g,q\}}(x, \xi ; \mu _{F}, \mu _{R})\,,
\end{align} 
in the region where $ 1>|x| >\xi $, which corresponds to the so-called DGLAP region of the GPD.  For $ \mu_{F} \sim 3 \GeV $, relevant for $ J/\psi $, the gluon GPD is rather flat in $ x $. Upon integration, the coefficient function in \EQ\eqref{eq:asy-amp} gives rise to a \textit{purely imaginary} amplitude enhanced by $ \ln(1/\xi) $ at small $ \xi $. Being proportional to $ \ln (m_{Q}^2/\mu^{2}_{F}) $, this leads to huge $ \mu_{F} $-scale uncertainties, shown in blue on the left panel of \FIG\ref{fig:jpsi-uncertainty} for $ J/\psi $. We highlight that the same effect is observed for the NLO quark contribution, since the quark singlet GPD scales as $ 1/x $ at small $ \xi $.

Large scale uncertainties are a sign of important higher order corrections. Indeed, the appearance of the large logarithms $ \ln (1/\xi) $ indicates that corrections of the type $ (\alpha_s \ln (1/\xi))^n$ need to be resummed. In fact, such a resummation through high-energy factorisation has been adopted for the
inclusive quarkonium hadro- and photoproduction, see \REF\cite{Lansberg:2021vie,Lansberg:2023kzf}.

\section{High-energy resummation and matching}

The large logarithms of $ \ln (1/\xi) $ observed in the amplitude actually have their origin at the integrand level, in the coefficient function $ C_{i}(x,\xi;\mu_{F},\mu_{R}) $, which develops a series of corrections $ (\alpha_s^{n}\ln^{n-1}(1/|\rho|))/|\rho| $, where $ \rho = \xi/x $. They can be resummed in the leading-logarithmic approximation (LLA) using the high-energy factorisation (HEF) formalism. Such a resummation has been originally developed for inclusive processes, but through the optical theorem, it can also be applied to exclusive reactions.

An example diagram that contributes to the imaginary part of the amplitude in the LLA is shown on the right panel of \FIG\ref{fig:dvmp}. The thick helices correspond to the so-called \textit{Reggeised gluons}, and they connect gluon emissions that are strongly ordered in $ k^+ $. Note that the part of the cut diagram below the blue dashed-dotted horizontal line is the usual inclusive BFKL ladder. Its computation leads to the universal HEF resummation factor $ {\cal C}_{gi}( \rho ,\mathbf{k}_{\perp}^{2};\mu_{F}, \mu _{R}) $. Denoting the process-dependent part of the cut diagram above the horizontal blue line by $ h(\mathbf{k}_{\perp}^{2}) $, the coefficient function in HEF, $ C_{i}^{( \mathrm{HEF} )} $, which replaces $ C_i $ in \EQ\eqref{eq:amp}, is given by
\begin{align}
	\label{eq:HEF-coefficient-function}
	C_{i}^{ (\mathrm{HEF} )}( \rho ; \mu _{F}, \mu _{R}) = \frac{-i}{| \rho |}\int_{0}^{\infty}d\mathbf{k}_{\perp}^{2}\,{\cal C}_{gi}( \rho ,\mathbf{k}_{\perp}^{2};\mu_{F}, \mu _{R})h(\mathbf{k}_{\perp}^{2})\,,
\end{align}
as was first discussed in \REF\cite{Ivanov:2007je}. The details of the calculations can be found in \REF\cite{Flett:2024htj}. At this point, we highlight that we truncate the HEF resummation factor $ {\cal C}_{gi} $ in the LLA down to the doubly-logarithmic approximation (DLA). This is because $ {\cal C}_{gi} $ in the LLA generates $  \mu _{F} $-dependent  terms which will not be compensated by the fixed-order evolution of GPDs. In this way, only corrections of the type $  (\alpha _{s}^{n} \ln^{n-1} (1/ |\rho |)\ln^{n-1} ( \mu _{F}^{2}/\mathbf{k}_{\perp}^{2}))/| \rho | $ are resummed. 

We match the CF coefficient function in \EQ\eqref{eq:amp}, which has been calculated to NLO in $  \alpha _{s} $, with the one calculated in the DLA of HEF in \EQ\eqref{eq:HEF-coefficient-function} through a \textit{subtractive matching} procedure:
\begin{align}
	C_{i}^{( \mathrm{match.} )}(x, \xi ) = C_{i}^{(0)}(x, \xi ) + C_{i}^{(1)}(x, \xi ) +  \left[ \hat{C}_{i}^{( \mathrm{HEF} )}( \xi /|x|) -C_{i}^{(1,\, \mathrm{asy.} )}(x, \xi ) \right]  \theta (|x| - \xi)\,.
\end{align}
The $  \theta (|x|- \xi ) $ ensures that the resummation is applied only in the DGLAP region, see \EQ\eqref{eq:asy-amp}. Furthermore, the hat ( $ \hat{} $ ) on $ \hat{C}_{i}^{( \mathrm{HEF} )}( \xi /|x|) $ implies that the LO $  \alpha _{s}^{0} $ term has been subtracted from it. When $ 1>|x| \gg \xi  $ (i.e.~when $  \rho \ll 1 $), $ C_{i}^{(1)}(x, \xi ) \to C_{i}^{(1,\, \mathrm{asy.} )}(x, \xi ) $, by definition, see \EQ\eqref{eq:asy-amp}, which means that the resummation is important in this region, since there is a huge phase space available for gluon emissions strongly ordered in $ k^+ $ (see right panel of \FIG\ref{fig:dvmp}). On the other hand, when $|x| \sim |\xi|$ (i.e.~when $  \rho \sim 1 $), it can be shown that $ \hat{C}_{i}^{( \mathrm{HEF} )}( \xi /|x|) \to C_{i}^{(1,\, \mathrm{asy.} )}(x, \xi )  $ (since $ \ln ^{n}(1/| \rho |) \to 0$), such that $ C_{i}^{ (\mathrm{match.}) }(x, \xi ) $ basically becomes the fixed order result in CF.

\section{Results}

For the GPD input to our calculations, we use the Goloskokov-Kroll (GK) model \cite{Goloskokov:2006hr},  which is itself based on the standard double distribution (DD) model of GPDs \cite{Radyushkin:1997ki}. Furthermore, we include a full leading-log (LL) evolution of GPDs, with the 1-loop kernels in \REF\cite{Bertone:2022frx}, using PARTONS \cite{Berthou:2015oaw}, interfaced to the APFEL++ code \cite{Bertone:2017gds}. We start the evolution at an initial scale of $  \mu _{0} = 2 \GeV $.

In \FIG\ref{fig:jpsi-uncertainty}, the relative scale uncertainty for $ J/\psi $ photoproduction with respect to  $ \mu_{F}  $ ($  \mu _R $) is shown on the left (right) panel, as a function of the centre-of-mass energy $ W_{\gamma p} $. The uncertainty consists of the ratio of the differential cross section in $ t $ evaluated at $ t = t_{ \mathrm{min} } $, $ \frac{d \sigma }{dt }\big{|}_{t_{ \mathrm{min} }}$ , varying the scales by a factor of 2. $ t_{ \mathrm{min} } $ is the lowest kinematically allowed value of $ t  $, and is basically equal to zero at high energies. We focus on the CT18NLO PDF set as the input for the GK DD model for our GPDs. As mentioned in \SEC\ref{sec:amp-coll-fact}, one finds that the fixed order NLO CF result (blue) gets huge scale uncertainties at high energies. The matched result of CF and HEF in the DLA, shown in red in \FIG\ref{fig:jpsi-uncertainty} remains stable at high energies, and has smaller uncertainties than the LO CF result when $  \mu _{F} $ is varied. For the $ \mu_R $ variation, one observes that the uncertainty for the matched result NLO CF $ \oplus  $ DLA HEF slowly grows at high energies, becoming even larger than the LO CF result. This is a consequence of the $  \mu _{R} $ dependence of the hard pomeron contribution, which we expect to be suppressed in a complete NLLA calculation.

\begin{figure}
\includegraphics[width=5.4cm, angle=-90]{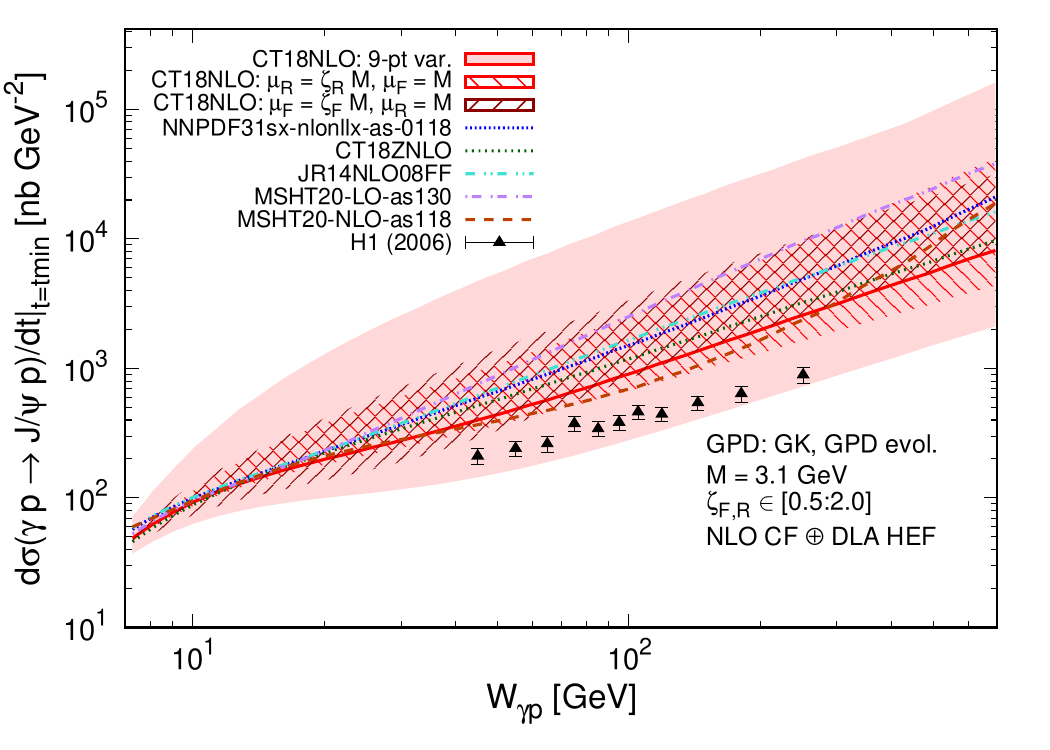}
\hfill
\includegraphics[width=5.4cm, angle=-90]{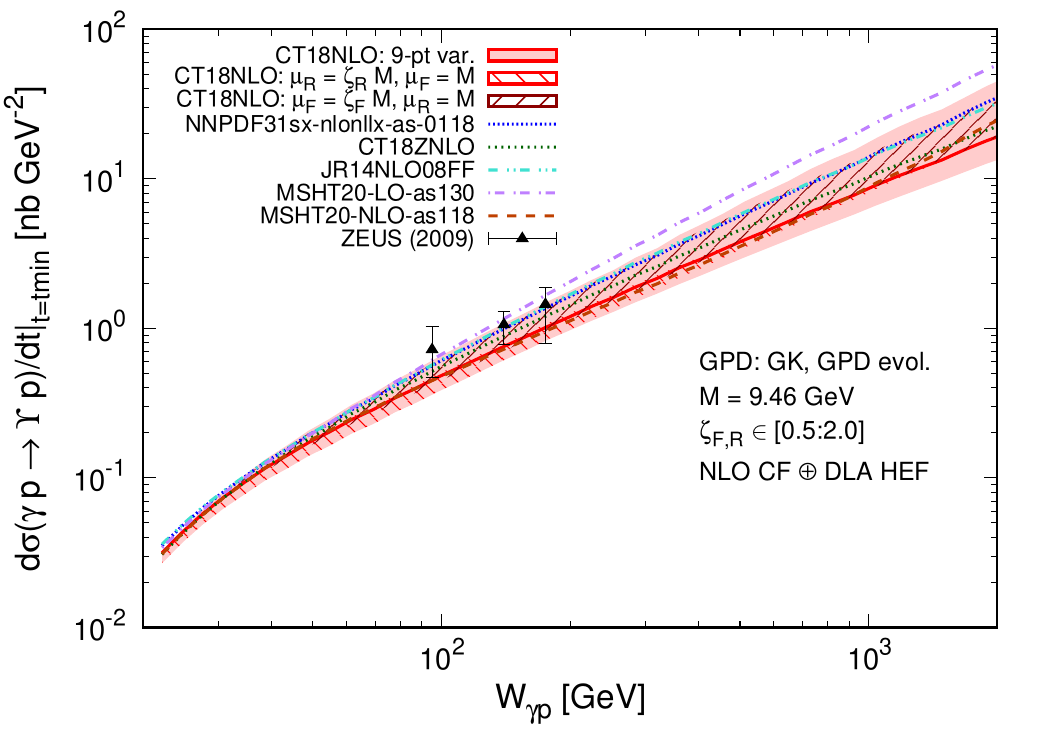}
\caption{The 9-point scale variation of the $ t $-differential cross section for $ J/\psi $ ($ \Upsilon $) as a function of the centre-of-mass energy $ W_{\gamma p} $ using the matched NLO CF $ \oplus $ DLA HEF results on the left (right).}
\label{fig:jpsi-upsilon-cross-section}
\end{figure}

In \FIG\ref{fig:jpsi-upsilon-cross-section}, the $ t $-differential cross section evaluated at $ t=t_{ \mathrm{min} } $ is shown as a function of $ W_{\gamma p} $ for $ J/\psi $ and $ \Upsilon $ on the left and right panels respectively, using the matched NLO CF $ \oplus $ DLA HEF result. A 9-point scale variation is performed for the GPDs constructed from the CT18NLO PDF set. Moreover, the central value of the cross sections for GPDs constructed using 5 different PDF sets are also shown in order to roughly assess the PDF uncertainty. For $ J/\psi $, it is observed that the precise H1 data from HERA \cite{H1:2013okq} is in line with our predictions, although the central values overshoot the experimental values. On the other hand, our predictions for $ \Upsilon  $ production, which have much smaller scale uncertainties compared to $ J/\psi $, are completely compatible with the ZEUS data \cite{ZEUS:2009asc}. The difference in compatibility between the two heavy vector mesons could be the consequence of important relativistic-$ v^2 $ and/or higher twist corrections, which are of course expected to be larger for $ J/\psi $ than $ \Upsilon $, or could come from a crude modelling of the gluon GPDs.

\bibliographystyle{utphys}

\bibliography{masterrefs.bib}

\end{document}